\documentclass[12pt]{article}

\usepackage{array,dsfont} 
\usepackage{epsfig}
\usepackage{amssymb}
\usepackage{graphics,graphpap}

\renewcommand{\thefootnote}{\fnsymbol{footnote}}

\newcommand{\lsim}{
\mathrel{\hbox{\rlap{\hbox{\lower4pt\hbox{$\sim$}}}\hbox{$<$}}}}

\newcommand{\gsim}{
\mathrel{\hbox{\rlap{\hbox{\lower4pt\hbox{$\sim$}}}\hbox{$>$}}}}

\begin{document}

\begin{titlepage}
\begin{flushright}\begin{tabular}{l}
IPPP/06/95\\
DCPT/06/190
\end{tabular}
\end{flushright}
\vskip1.5cm
\begin{center}
   {\Large \bf\boldmath Constraints on New Physics from $\gamma$ and
   $|V_{ub}|$}
    \vskip2.5cm {\sc
Patricia Ball\footnote{Patricia.Ball@durham.ac.uk}
}
  \vskip0.5cm
{\em         IPPP, Department of Physics,
University of Durham, Durham DH1 3LE, UK }\\
\vskip2.5cm 


\vskip2.5cm

{\large\bf Abstract\\[10pt]} \parbox[t]{\textwidth}{
The SM unitarity triangle (UT) is completely determined by the parameters 
$\gamma$ and $|V_{ub}|$ which can be extracted from tree-level
processes and are assumed to be free of new physics. By comparison
with other determinations of UT parameters one can 
impose constraints on new physics in loop processes, in particular $B$ mixing.
}
\vskip1cm

$$\epsfxsize=0.4\textwidth\epsffile{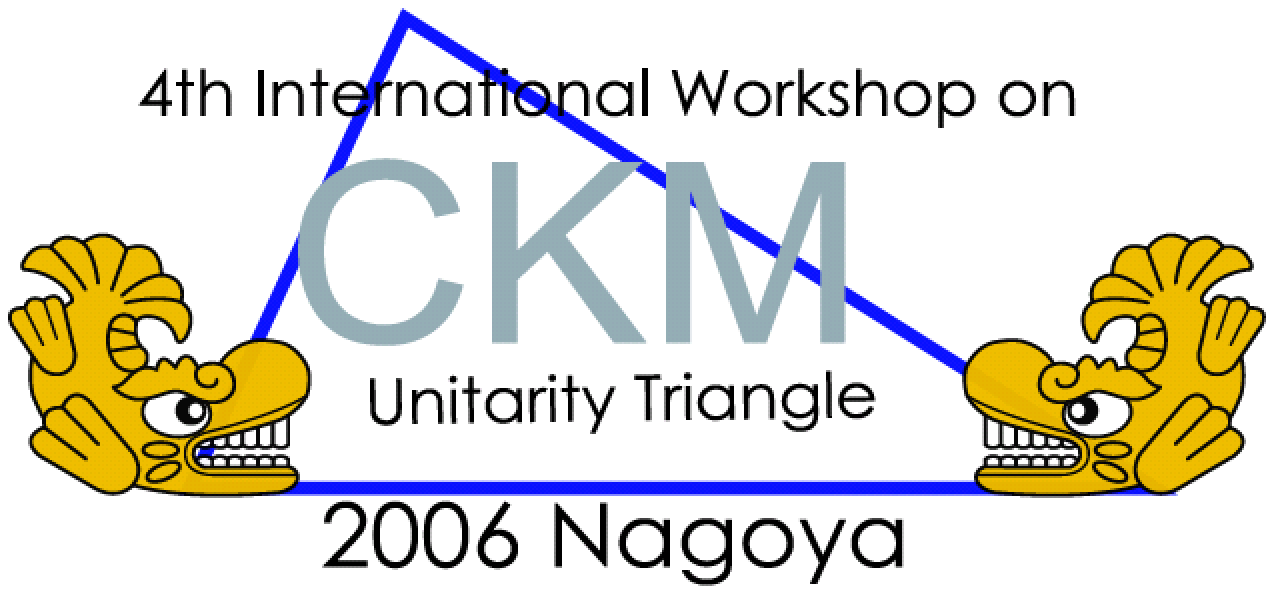}$$

\centerline{\em Talk given at CKM06, Nagoya (Japan), Dec 2006}
\end{center}

\vfill

\end{titlepage}

\setcounter{footnote}{0}
\renewcommand{\thefootnote}{\arabic{footnote}}

\newpage

Independently of any new sources of flavour violation induced by new
physics (NP), there is always a
Standard Model (SM) unitarity triangle (UT). It is completely
determined by two parameters, which one can choose as
$|V_{ub}/V_{cb}|$ and $\gamma$ -- the rationale being that these
parameters can be determined from tree-level processes and hence are
expected to be essentially free of new-physics effects. In this talk
we discuss the impact of the presently available information on
$|V_{ub}/V_{cb}|$ and $\gamma$ on possible new physics in $B$
mixing, based on Ref.~\cite{PB4}; we include the most recent update on 
$(\sin\phi_d)_{c\bar c s}$ presented at ICHEP2006. 

Let us first discuss the status of $|V_{ub}|$ and $|V_{cb}|$. 
The latter quantity is presently known with 
2\% precision from semileptonic $B$ decays;  we shall use the 
value obtained in Ref.~\cite{Buchmuller} from the analysis of leptonic and 
hadronic moments in inclusive $b\to c \ell \bar\nu_\ell$ transitions 
\cite{Gambino}:
\begin{equation}\label{Vcb}
|V_{cb}| = (42.0\pm 0.7)\cdot 10^{-3}\,;
\end{equation}
this value agrees with that from exclusive semileptonic decays. 

The situation is
less favourable with $|V_{ub}|$: there is more than 1$\sigma$ discrepancy
between the values from inclusive and exclusive $b\to u\ell\bar\nu_\ell$
transitions \cite{HFAG}:
\begin{equation}
|V_{ub}|_{\rm incl} = (4.4\pm 0.3)\cdot 10^{-3}\,,\qquad 
|V_{ub}|_{\rm excl} = (3.8\pm 0.6)\cdot 10^{-3}\,.
\end{equation}
The error on $|V_{ub}|_{\rm excl}$ is dominated by the theoretical
uncertainty of lattice and light-cone sum rule calculations of $B\to\pi$ and
$B\to\rho$ transition form factors \cite{Vublatt,LCSR}, whereas for
$|V_{ub}|_{\rm incl}$ experimental and theoretical errors are at
par. A recent improvement of the method used to extract $|V_{ub}|$ has 
been suggested in Ref.~\cite{PB7}; it relies on
 fixing the shape of the exclusive
form factor from experimental data on the $q^2$-spectrum in $B\to\pi e
\nu$, which helps to reduce both the experimental and theoretical
error of $|V_{ub}|_{\rm excl}$. 
The ``low'' value $|V_{ub}|_{\rm excl}$ is in agreement with the
determination of $|V_{ub}|$, by the UTfit collaboration, 
from only the angles of the UT \cite{UTfit}. 
In this report we shall present results
for both values of $|V_{ub}|$.

As for the UT angle $\gamma$, tree-level results can be obtained from
the CP asymmetries in $B\to D^{(*)} K^{(*)}$ decays. At present, the
only results available come from the  
Dalitz-plot analysis of the CP asymmetry in $B^-\to (K_S^0 \pi^+\pi^-)
K^-$, with $K_S^0 \pi^+\pi^-$ being a three-body final state common
to both $D^0$ and $\bar D^0$. This method to measure $\gamma$ from a
new-physics free tree-level process was suggested in Ref.~\cite{GGSZ}
and has been implemented by both BaBar \cite{babargamma} and Belle 
\cite{Bellegamma}, but the BaBar result currently suffers from huge
errors: $\gamma_{\rm BaBar} = (92\pm 41\pm 11\pm 12)^\circ$,
$\gamma_{\rm Belle} = (53^{+15}_{-18}\pm 3\pm 9)^\circ$.
Other determinations of $\gamma$ from QCDF, $\gamma_{\rm QCDF} =
(62\pm 8)^\circ$ \cite{MN}, SCET, $\gamma_{\rm SCET} =
(73.9^{+7.4}_{-10.7})^\circ$ 
\cite{is}, SU(3) fits of non-leptonic $B$ decays
$\gamma_{\rm SU(3)} = (70.0^{+3.8}_{-4.3})^\circ$
\cite{rf}, radiative penguin decays,
$\gamma_{B\to V\gamma} = (61.0^{+13.5}_{-16.0}{}^{+8.9}_{-9.3})^\circ$
\cite{radpeng}, and global UT fits \cite{UTfit,CKMfitter}
all come with theoretical uncertainties and/or possible contamination by
unresolved new physics. In this report we shall use $\gamma=(65\pm
20)^\circ$, which is a fair average over all these determinations.

With $\gamma$ and $|V_{ub}/V_{cb}|$ fixed,
let us first have a closer look at the $B^0_d$--$\bar B^0_d$ mixing
parameters.
In the presence of NP, the matrix element $M_{12}^d$ can be
written, in a model-independent way, as
$$M_{12}^d = M_{12}^{d,{\rm SM}} \left(1 + \kappa_d e^{i\sigma_d}\right)\,,$$
where the real parameter $\kappa_d\geq 0$ measures the ``strength'' of
the NP contribution with respect to the SM, whereas $\sigma_d$ is a new CP-violating
phase; analogous formulae apply to the $B_s$ system. 
The $B_d$ mixing parameters then read
\begin{eqnarray}
\Delta M_d & = & \Delta M_d^{\rm SM}\left[ 1 + \kappa_d
  e^{i\sigma_d}\right],\label{M-d}\\
\phi_d & = & \phi_d^{\rm SM}+\phi_d^{\rm NP}=
\phi_d^{\rm SM} + \arg (1+\kappa_d e^{i\sigma_d})\,.\label{phi-d}
\end{eqnarray}

Experimental constraints on $\kappa_d$ and $\sigma_d$ are provided by
$\Delta M_d$ and $\phi_d$, the mass difference and mixing phase in the
$B_d$ system. While the interpretation of the very accurately known
experimental value of $\Delta M_d$ depends crucially on hadronic
matrix elements provided by lattice calculations, $\phi_d$ can be
measured directly as mixing-induced CP asymmetry in $b\to c\bar c s$
transitions \cite{HFAG}:
\begin{equation}
(\sin\phi_d)_{c\bar s} = 0.675\pm 0.026\,,
\end{equation}
which yields the twofold solution
\begin{equation}\label{phid-det}
\phi_d=(42.5\pm2.0)^\circ \quad\lor\quad (137.5\pm2.0)^\circ,
\end{equation}
where the latter result is in dramatic conflict with global CKM fits and
would require a large NP contribution to $B^0_d$--$\bar B^0_d$
mixing. However, experimental information on the sign of $\cos\phi_d$ 
rules out a negative value of this quantity at greater than 95\% C.L.\
\cite{WG5-report}, so that we are left with $\phi_d=(42.5\pm2.0)^\circ$.

The SM prediction of the mixing phase, $\phi_d^{\rm SM}=2\beta$,
can easily be obtained in terms of the tree-level quantities 
$R_b$ and $\gamma$, as
\begin{equation}\label{beta-true}
\sin\beta=\frac{R_b\sin\gamma}{\sqrt{1-2R_b\cos\gamma+R_b^2}}\,, \quad
\cos\beta=\frac{1-R_b\cos\gamma}{\sqrt{1-2R_b\cos\gamma+R_b^2}}\,.
\end{equation}
Here the quantity $R_b$ is given by
\begin{equation}\label{Rb-def}
R_b\equiv\left(1-\frac{\lambda^2}{2}\right)\frac{1}{\lambda}
\left|\frac{V_{ub}}{V_{cb
}}\right|.
\end{equation}
Using Eq.~(\ref{phi-d}), the experimental value of $\phi_d$ can 
immediately be converted into a result for the NP phase $\phi_d^{\rm
  NP}$, which depends on both $\gamma$ and $R_b$.
It turns out that the dependence of
$\phi^{\rm NP}_d$ on $\gamma$ is very small and that $R_b$ plays
actually the key r\^ole for its determination.
With our range of values for $\gamma$ and $|V_{ub}|$ we find
\begin{equation}
\left.\phi_d^{\rm SM}\right|_{\rm incl} = 
(53.5\pm 3.8)^{\circ}\,,\qquad \left.\phi_d^{\rm SM}\right|_{\rm excl} =
(45.9\pm 7.6)^{\circ}\,,
\end{equation}
corresponding to 
\begin{equation}\label{phiNPd-num}
\left.\phi^{\rm NP}_d\right|_{\rm incl} = -(11.0\pm 4.3)^\circ\,,\qquad
\left.\phi^{\rm NP}_d\right|_{\rm excl} = -(3.4\pm 7.9)^\circ\,;
\end{equation}
results of $\phi_d^{\rm NP}\approx-10^\circ$ were also recently obtained in 
Refs.~\cite{BFRS-05,UTfit-NP,blanke}. 
Note that the emergence of a  non-zero value of
$\phi_d^{\rm NP}$ is caused by the large value of $|V_{ub}|$ from
inclusive semileptonic decays, but that $\phi_d^{\rm NP}$ is
compatible with zero for $|V_{ub}|$ from exclusive decays.

We can now combine the constraints from both
$\Delta M_d$ and $\phi_d$ to constrain the allowed region in the 
$\sigma_d$--$\kappa_d$ plane. These contraints depend on hadronic
input for $\Delta M_d$ in terms of the parameter
$f_{B_d}\hat{B}_{B_d}^{1/2}$ for which there exist two independent
unquenched lattice results, one by the JLQCD collaboration with
$N_f=2$ active flavours \cite{JLQCD}, and one by the HPQCD
collaboration with $N_f=2+1$ active flavours \cite{HPQCD}. We also
give the corresponding results for the $B_s$ which we will need below:
\begin{eqnarray}
\left.f_{B_d}\hat{B}_{B_d}^{1/2}\right|_{\rm JLQCD} &=& (0.215\pm
0.019^{+0}_{-0.023})\,{\rm GeV}\,,\nonumber\\
\left.f_{B_s}\hat{B}_{B_s}^{1/2}\right|_{\rm JLQCD} &=& (0.245\pm
0.021^{+0.003}_{-0.002})\,{\rm GeV}\,,\label{JLQCD}\\
\left.f_{B_d}\hat{B}_{B_d}^{1/2}\right|_{\rm (HP+JL)QCD}& =& (0.244\pm
0.026)\,{\rm GeV}\,,\nonumber\\
\left.f_{B_s}\hat{B}_{B_s}^{1/2}\right|_{\rm HPQCD} &=&
(0.281\pm0.021)\,{\rm GeV}\,.\label{HPQCD}
\end{eqnarray}
The last but one entry is a combination of both HPQCD and JLQCD
results, as the HPQCD collaboration is yet to provide results on $B_{B_d}$.

\begin{figure}[t]
$$\epsfxsize=0.47\textwidth\epsffile{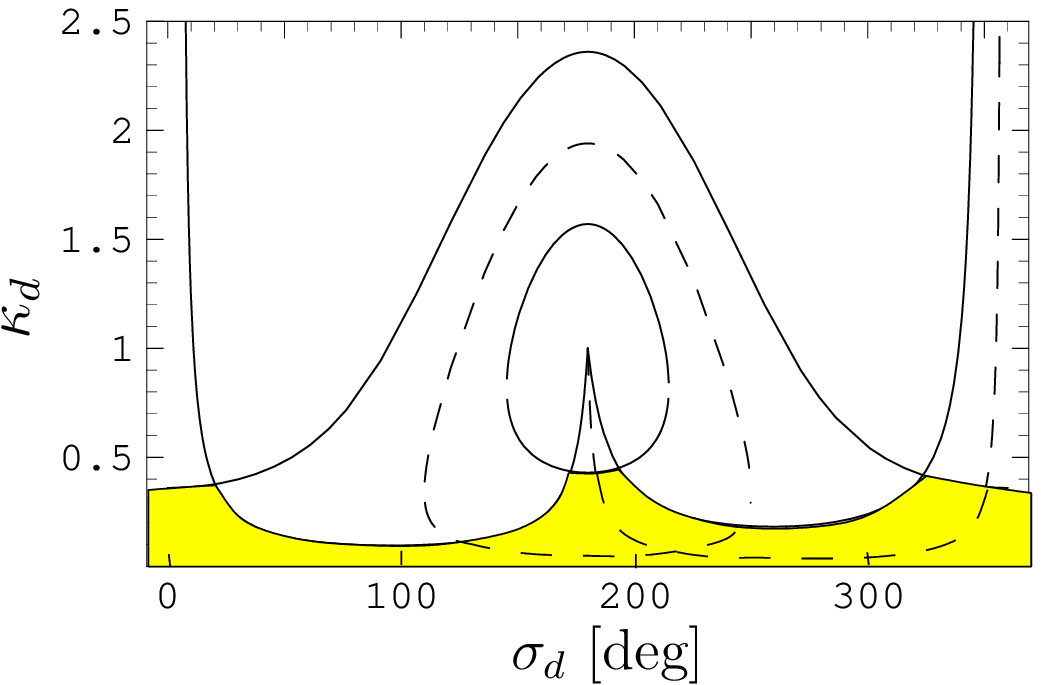}\quad
\epsfxsize=0.47\textwidth\epsffile{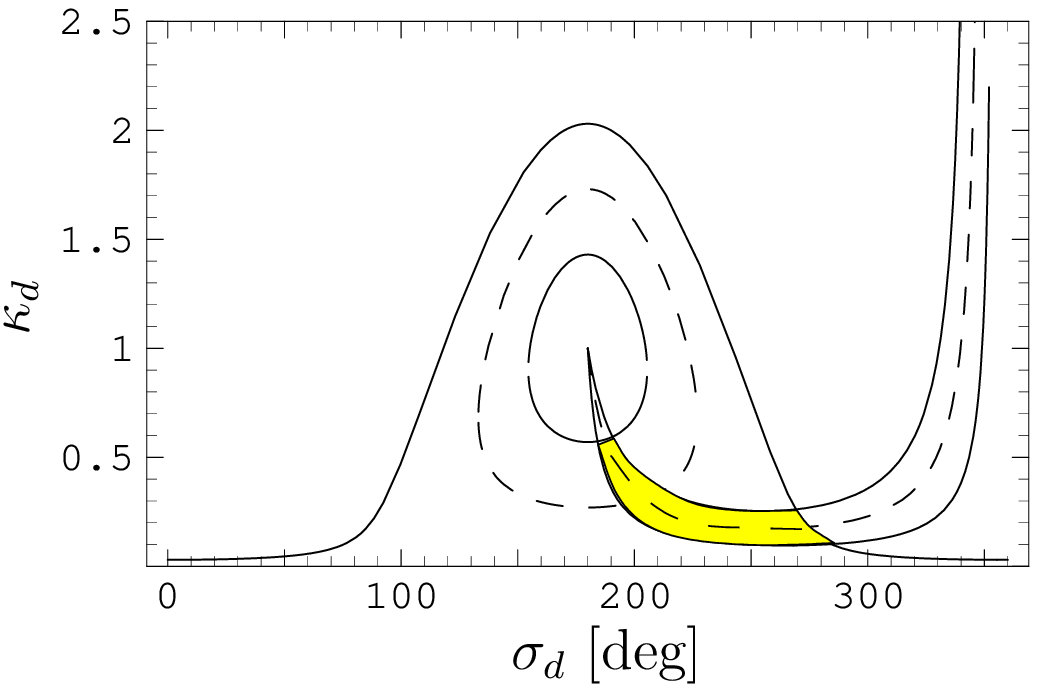}
$$
 \vspace*{-1truecm}
\caption[]{\small Left panel: allowed region (yellow/grey) in the 
$\sigma_d$--$\kappa_d$
  plane in a scenario with the JLQCD lattice results (\ref{JLQCD}) and 
  $\left.\phi^{\rm NP}_d\right|_{\rm excl}$. Dashed lines: central
  values of $\Delta M_d^{\rm latt}$ 
  and $\phi^{\rm NP}_d$, solid lines: $\pm 1\,\sigma$. Right panel: ditto for the 
 scenario with the (HP+JL)QCD   lattice results
  (\ref{HPQCD}) and  $\left.\phi^{\rm NP}_d\right|_{\rm incl}$. 
}\label{fig:res-k-sig-d}
\end{figure}

The corresponding constraints in the $\sigma_d$-$\kappa_d$ plane are shown
in Fig.~\ref{fig:res-k-sig-d}. We see that a non-vanishing value
of $\phi_d^{\rm NP}$, even as small as $\phi_d^{\rm NP}\approx -10^\circ$,
has a strong impact on the allowed space in the $\sigma_d$--$\kappa_d$ plane.
In both scenarios with different lattice results and different
values for $|V_{ub}|$, the upper bounds of $\kappa_d\lsim2.5$ on the NP
contributions following from the experimental value of $\Delta M_d$ are reduced
to $\kappa_d\lsim0.5$. 
In order to determine $\kappa_d$ more precisely, it is mandatory
to reduce the errors of $\Delta M_d^{\rm latt}$,  which come from both $\gamma$ and
lattice calculations. The value of $\gamma$ can be
determined -- with impressive accuracy -- at the LHC, whereas
progress on the lattice side is much harder to predict.

\begin{figure}[t]
$$\epsfxsize=0.47\textwidth\epsffile{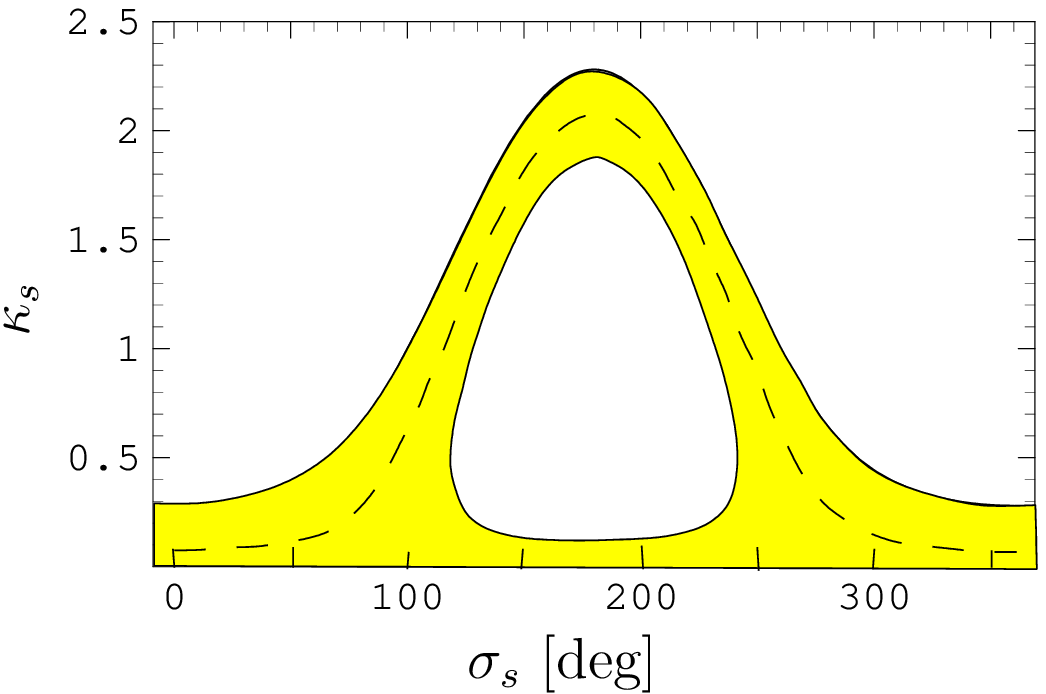}\quad
\epsfxsize=0.47\textwidth\epsffile{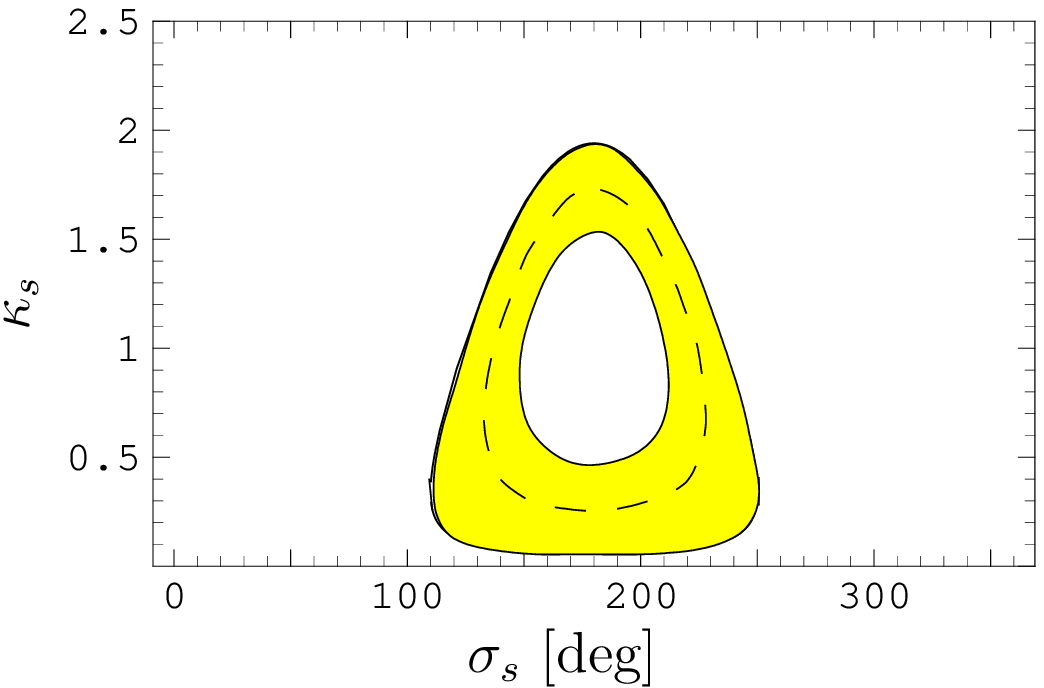}
$$
 \vspace*{-1truecm}
\caption[]{\small The allowed regions (yellow/grey) in the $\sigma_s$--$\kappa_s$ plane.
Left panel: JLQCD lattice results (\ref{JLQCD}). Right panel: HPQCD lattice 
results (\ref{HPQCD}).}\label{fig:MDs-NP}
\end{figure}

Let us now have a closer look at the $B_s$-meson system. The big news
in 2006 was the first measurement, by the CDF collaboration, of
$\Delta M_s$ \cite{deltams}: 
\begin{equation}
\Delta M_s = (17.77\pm 0.10\pm 0.07)\,{\rm ps}^{-1}\,.
\end{equation}
In order to describe
NP effects in $B_s$ mixing 
in a model-independent way, we parametrize them analogously
to (\ref{M-d}) and (\ref{phi-d}). The relevant CKM factor is
$|V_{ts}^* V_{tb}|$. Using the
unitarity of the CKM matrix and including next-to-leading
order terms in the Wolfenstein expansion, we have
\begin{equation}\label{Vts}
\left|\frac{V_{ts}}{V_{cb}}\right|=1-\frac{1}{2}\left(1-2R_b\cos\gamma\right)\lambda^2
+{\cal O}(\lambda^4).
\end{equation}
Consequently, apart from the tiny correction in $\lambda^2$, the  CKM
factor for $\Delta M_s$ is independent of $\gamma$ and $R_b$,
which is an important advantage in comparison with the $B_d$-meson system.
The accuracy of the SM prediction of $\Delta M_s$ is hence  limited by the
hadronic mixing parameter $f_{B_s}\hat{B}_{B_s}^{1/2}$.
In Fig.~\ref{fig:MDs-NP}, we show the constraints in the $\sigma_s$--$\kappa_s$ 
plane. We see that  upper bounds of $\kappa_s\lsim 2.5$
arise from the measurement of $\Delta M_s$. Consequently, 
the CDF measurement of $\Delta M_s$ leaves ample space for the 
NP parameters $\sigma_s$ and $\kappa_s$. This situation will change significantly 
as soon as precise information about CP violation in the $B_s$-meson system becomes
available.

To date, the CP-violating phase associated with $B^0_s$--$\bar B^0_s$ mixing 
is not very well constrained. In the SM, it is doubly Cabibbo-suppressed, and 
can be written as follows:
\begin{equation}\label{phis-SM}
\phi_s^{\rm SM}=-2\lambda^2\eta=
-2\lambda^2R_b\sin\gamma \approx -2^\circ.
\end{equation}
Because of the
small SM phase in (\ref{phis-SM}), $B^0_s$--$\bar B^0_s$ mixing
is particularly well suited to search for NP effects, which may well
lead to a sizeable value of $\phi_s$. The presently
available information on $\phi_s$ stems from measurements of
$\Delta\Gamma_s$ and the semileptonic CP asymmetry $a_{fs}^s$; they
have been re-analysed very recently in Ref.~\cite{LenzNierste} with
the result
\begin{equation}
\sin \phi_s = -0.77\pm 0.04\pm 0.34 \quad \mbox{or}\quad 
\sin\phi_s = -0.67\pm 0.05\pm 0.29\,,
\end{equation}
depending on the value of $\Delta M_s^{\rm latt}$; both results would imply a
 2$\sigma$ deviation from the SM prediction $\sin\phi_s^{\rm SM}
 \approx -0.04$, but are heavily theory dependent.
 In order to test
the SM and  probe CP-violating NP contributions to 
$B^0_s$--$\bar B^0_s$ mixing in a less theory-dependent way, 
the decay $B^0_s\to J/\psi\phi$, which
is very accessible at the LHC,  plays a key r\^ole
and allows the measurement of 
\begin{equation}
\sin\phi_s=\sin(-2\lambda^2R_b\sin\gamma+\phi_s^{\rm NP})\,, 
\end{equation}
in analogy to the determination of $\sin\phi_d$
through $B^0_d\to J/\psi K_{\rm S}$. 

\begin{figure}[t] 
$$\epsfxsize=0.47\textwidth\epsffile{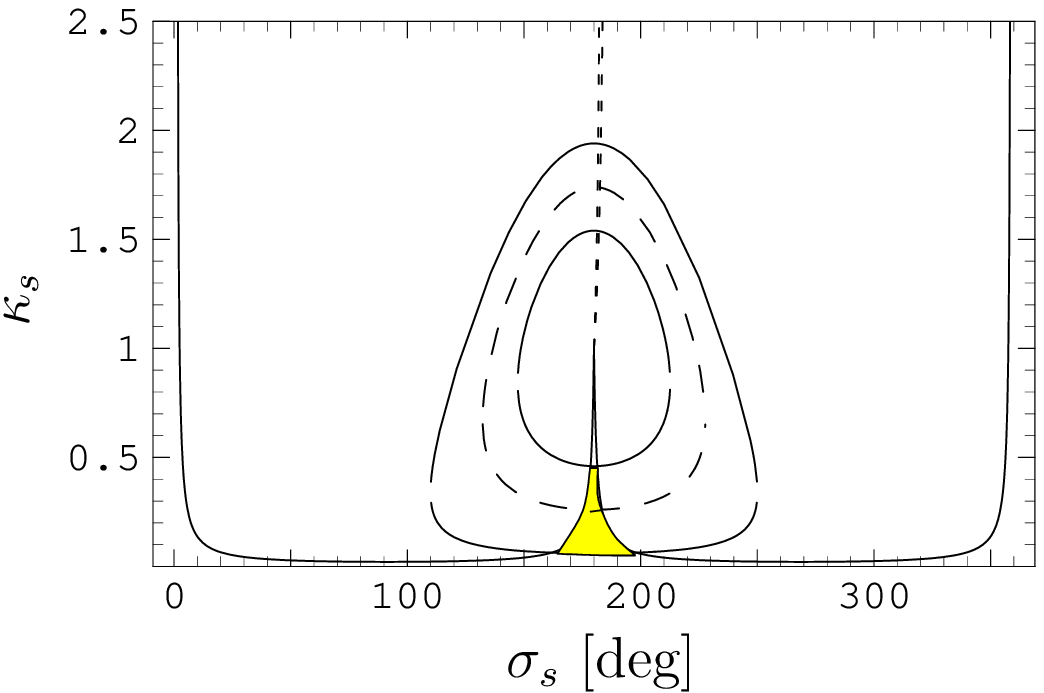}\quad
\epsfxsize=0.47\textwidth\epsffile{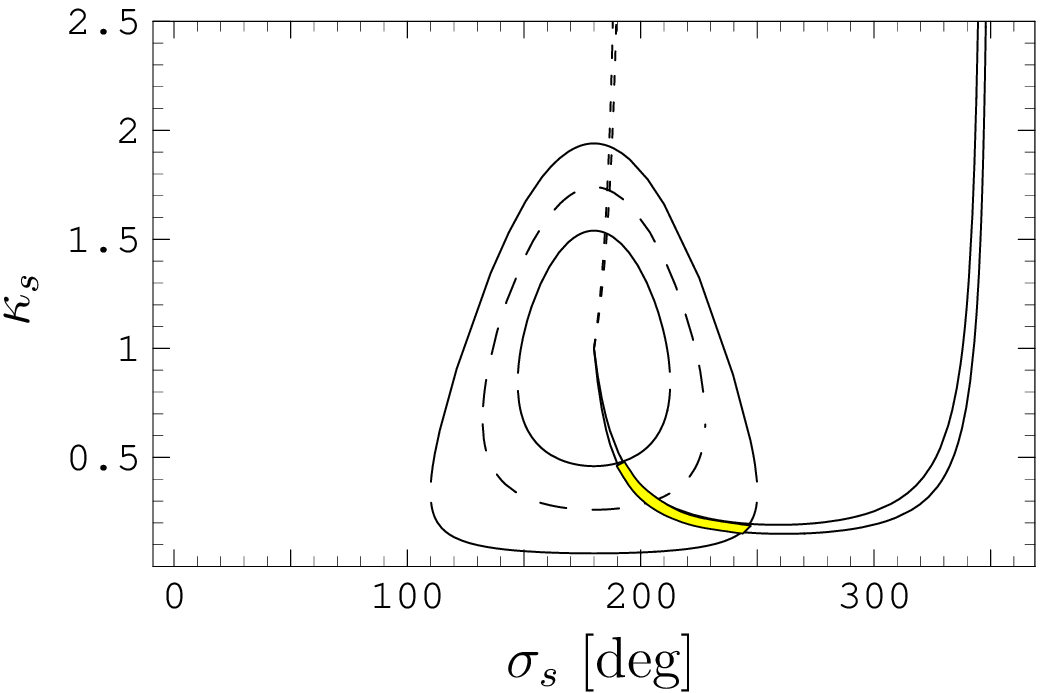}$$
\vspace*{-1cm}
   \caption[]{\small Combined constraints for the allowed region (yellow/grey) in the 
   $\sigma_s$--$\kappa_s$ plane through $\Delta M_s$
   for the HPQCD results (\ref{HPQCD}) and CP violation measurements.
   Left panel: the SM scenario $(\sin\phi_s)_{\rm exp}=-0.04\pm0.02$. Right panel: 
   a NP scenario with $(\sin\phi_s)_{\rm exp}=-0.20\pm0.02$. The solid
   lines correspond to $\cos\phi_s>0$, the dotted
   lines to  $\cos\phi_s<0$.}\label{fig:sis-kas-CP}
\end{figure}

In order to illustrate the possible impact of NP effects, let us assume that
the NP parameters satisfy the simple relation
\begin{equation}\label{sig-kap-rel}
\sigma_d=\sigma_s,  \quad \kappa_d=\kappa_s, 
\end{equation}
i.e.\ that in particular $\phi_d^{\rm NP}=\phi_s^{\rm NP}$. 
To illustrate the impact of CP violation measurements 
on the allowed region in the $\sigma_s$--$\kappa_s$ plane, 
let us consider two cases:
\begin{itemize}
\item[i)] $(\sin\phi_s)_{\rm exp}=-0.04\pm0.02$, i.e.\ the SM
  prediction;
\item[ii)] $(\sin\phi_s)_{\rm exp}=-0.20\pm0.02$, i.e.\   the above NP 
scenario $\phi_d=\phi_s\approx-11^\circ$.
\end{itemize}
In 
Fig.~\ref{fig:sis-kas-CP}, we show the situation in the
$\sigma_s$--$\kappa_s$ plane. The constraints on the NP parameters are
rather strong, although $\kappa_s$ could still
assume sizeable values, with the upper bound $\kappa_s\approx0.5$.
In the SM-like scenario (i), values of $\sigma_s$ around
$180^\circ$ would arise, i.e.\ a NP contribution with a sign opposite to 
the SM. However, due to the absence of new CP-violating effects, 
the accuracy of lattice results would have to be considerably improved
in order to allow the extraction of a value of  $\kappa_s$ incompatible with 0.
On the other hand, a measurement of $(\sin\phi_s)_{\rm exp}=-0.20\pm0.02$
would give a NP signal at the $10\,\sigma$ level, with
$\kappa_s\gsim0.2$. 

Let us conclude with a few remarks concerning the prospects for the search for NP through $B^0_s$--$\bar B^0_s$
mixing at the LHC. This task will be very challenging if essentially no CP-violating effects
will be found in $B^0_s\to J/\psi \phi$ (and similar decays). On the other hand,
as we demonstrated above, even a small phase 
$\phi_s^{\rm NP}\approx-10^\circ$ (inspired by the $B_d$ data) would lead
to CP asymmetries at the $-20\%$ level, which could be unambiguously detected 
after a couple of years of data taking, and would not be affected by
hadronic uncertainties. Conversely, the measurement of such an asymmetry would 
allow one to establish a lower bound on the strength of the NP contribution -- even if 
hadronic uncertainties still preclude a direct extraction of this contribution from 
$\Delta M_s$ -- and to dramatically reduce the allowed region in the NP parameter 
space. In fact, the situation may be even more promising, as specific scenarios of NP 
still allow large new phases in $B^0_s$--$\bar B^0_s$ mixing, also after the 
measurement of $\Delta M_s$, see, for instance, Refs.~\cite{emi,JMF}.

In essence, the lesson to be learnt from the CDF measurement of
$\Delta M_s$ is that NP may actually be hiding in $B^0_s$--$\bar B^0_s$
mixing, but is still obscured by parameter uncertainties, some of which will 
be reduced by improved statistics at the LHC, whereas others require dedicated 
work of, in particular, lattice theorists. The smoking gun for the presence of NP 
in $B^0_s$--$\bar B^0_s$ mixing will be the detection of a non-vanishing 
value of $\phi_s^{\rm NP}$ through CP violation in $B^0_s\to J/\psi\phi$. Let us
finally emphasize that the current $B$-factory data may show -- in addition 
to $\phi_d^{\rm NP}\approx-10^\circ$ -- other first indications of new sources of
CP violation through measurements of $B^0_d\to\phi K_{\rm S}$ and $B\to\pi K$
decays, which may point towards a modified electroweak penguin sector. 
All these examples are yet another demonstration that flavour physics is
not an optional extra, but an indispensable
ingredient in the pursuit of NP, also and in particular in the era of the LHC.

\end{document}